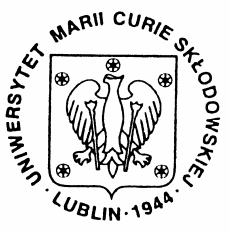

# Deterministic method of data sequence processing

Michał Widera*

*Institute of Medical Technology and Equipment ITAM, Roosevelt 118, 41-800 Zabrze, Poland*

**Abstract**

A data management system can be separated in typical data processing systems. Unfortunately, relational data management systems are not efficient enough to handle the on-line signal processing task in a monitoring system. The main current in research into database management system model for the needs of monitoring systems is connected with a data stream model. However, these systems are non-deterministic. This paper presents the developed methods of data stream processing for signal processing tasks in medical database management systems, as well as the developed theorems of data sequences (stream) algebra with formal proofs. A direct link between some introduced operators and Beatty and Fraenkel theorems has been proved.

## 1. Introduction

The currently available data management systems process data on the basis of relation calculus methods. This data processing method makes it possible to describe any objects (Entities) and the connection between them (Relation). The method of describing data structures by means of Entity-Relationship Diagrams [1] has become a binding standard in the design of data management system structures.

A data management system can be separated in typical data processing systems. The application of a relational database management system in the systems that feed data from external sources (e.g. a monitoring system) was not efficient enough [2]. The search for active or sequential database extensions [3] broadened the scope of knowledge in this field, but did not solve the problem. The established relational paradigm makes it difficult to increase efficiency [1].

The current scientific effort to create a data stream management system model for monitoring systems is connected with a data stream model and data management systems [4]. The application of such a system in a monitoring task will allow simple and elegant notations of algorithms. Moreover, owing to a simple system construction data processing will become much more efficient.

* *E-mail address*: michal@widera.int.pl

The main difference between the relational and stream systems lies in continuous queries [5]. By a continuous query we understand a query the realization plan of which is locked in an infinite loop. In a relational system, such a functionality is obtained by a relatively complicated and less effective method of continuous querying data (data base querying) by application or using several trigger sets. In a relational system, the data are introduced and approved by the user, on the other hand, in a stream system, the data are introduced from external, usually autonomic subsystems. Additionally, the data stream is infinite by nature. Traditional data management systems usually use data for which the maximum limit of resources has been estimated. Therefore, the use of infinite data sets requires the modification of some operators.

### 1.1. Data management systems

Medical monitoring systems record, analyze and present on-line information provided by medical devices [6]. The aim of research was to draw up a data management system and a query language that would enable forming and realization of signal processing tasks. The algebra was developed [7] and the query language was implemented in a data management system [8]. The developed solution enables effective signal processing for the needs of medical monitoring systems as well as simple realization and notation of parallel signal processing algorithms.

### 1.2. Projects dedicated to stream systems

Currently, there are a few leading research projects covering stream system issues [9,10]. The STREAM project [1] is being developed at Stanford University. The Aurora project, currently Borealis [11,12], is underway at MIT University. The authors of the STREAM project are looking for general rules of data stream processing based on the expansion of relation algebra. The definition of relation was modified for the needs of the developed data model. Relation $R$ is represented by a time dependent tuples set. The notation of the relation is presented in the form of $R(t)$. The existing operators were called relation-relational, while a set of relation-streaming and stream-relational operators was introduced for data stream processing. It is worth noting that there were no stream-streaming operators. In the Borealis project case the connection of relation and stream set is not openly considered. The introduced set of operators handles data stream operations only. This project has found commercial applications.

## 2. Stream algebras

Algebra is a branch of mathematics, while the term “general algebra” (or “algebra”) concerns a mathematical structure containing a data set and certain operations defined on this data set. The following mathematical structure can be an example of a simple definition of algebra: $A_1 := \left(N, \left(+, -\right)\right)$. Algebra $A_1$ was defined as a set of natural numbers that can by added and subtracted.

In the case of algebra based on data streams [11, 4], an ordered pair was used as a model. The first element of such a pair is a tuple (a tuple is a generalization of an ordered pair for unlimited count of elements) and the second is the time of occurrence – *(s, t)*.

### 2.1. Operations on data streams

When processing data streams, we assume that the following data exist:

- incoming data are in sequence,
- each element is checked and processed only once.

The main problem is to assure the required resource space to process a single element. In a perfect case, the requested size of memory and time of realization is independent of the size of data stream. The limited resources of the system prevent us from determining the exact query answer based on a data stream unlimited in time and dimension. In such a case, approximated answers are assumed to be satisfactory. Several methods of creating approximated algorithms are considered: drafts, sampling, histograms and wave algorithms. In the case of signal processing, a deterministic method must be applied. The presented assumptions of stream systems cannot be directly applied to solve such tasks.

## 3. Data sequence processing

The need for a deterministic method of data stream processing has been perceived in the course of research into real-time systems. Scientists working on the QStream project [13] have determined the assumptions for this type of data management systems. The presented results of research [7,14] go one step further. The deterministic algebra of data streams based on the number theory and covering systems has been proposed. The considered problem is the method of partitioning the set of positive natural numbers. Two sequences partition the set of positive natural numbers if the sequences are disjoint and their union is a set of natural numbers. The formal definition of the theorem providing a basis for the research was presented by S. Beatty in 1926 [15].

**Theorem 1 (Beatty).** *If p and q are positive irrationals such that* $1/p + 1/q = 1$ *then the sequences* $\left\{\lfloor np \rfloor\right\}_{n=1}^{\infty} = \lfloor p \rfloor, \lfloor 2p \rfloor, \lfloor 3p \rfloor, \ldots$ *and* $\left\{\lfloor nq \rfloor\right\}_{n=1}^{\infty} = \lfloor q \rfloor, \lfloor 2q \rfloor, \lfloor 3q \rfloor, \ldots$ *constitute a partition of the set of positive integers.*

The research in literature is based on the following sequence called Beatty sequence (floor).

$$\mathcal{B}(\alpha,\alpha') := \left( \left\lfloor \frac{n-\alpha'}{\alpha} \right\rfloor \right)_{n=1}^{\infty} \tag{1}$$

or Beatty sequence (ceiling):

$$\mathcal{B}^{(c)}(\alpha,\alpha') := \left( \left\lceil \frac{n-\alpha'}{\alpha} \right\rceil \right)_{n=1}^{\infty} . \tag{2}$$

Where $\alpha$ denotes density, $1/\alpha$ denotes slope, $\alpha'$ denotes offset and $-\alpha'/\alpha$ y-intercept.

The Beatty theorem makes it possible to partition the set with the help of two irrationals only. In 1969, Aviezri S. Fraenkel [16] generalized the previous theorem by presenting a method of creating sequences that partition this set with the help of two rational numbers.

**Theorem 2 (Fraenkel).** *The sequences* $\mathcal{B}(\alpha,\alpha')$ *and* $\mathcal{B}(\beta,\beta')$ *partition* $\mathbb{N}$ *if and only if the following five conditions are satisfied.*

1. $0<\alpha<1$.
2. $\alpha+\beta=1$.
3. $0\le\alpha+\alpha'\le1$.
4. *if* $\alpha$ *is irrational, then* $\alpha'+\beta'=0$ *and* $k\alpha+\alpha'\notin\mathbb{Z}$ *for* $2\le k\in\mathbb{N}$.
5. *if* $\alpha$ *is rational, (say* $q\in\mathbb{N}$ *is minimal with* $q\alpha\in\mathbb{N}$*), then* $1/q\le\alpha+\alpha'$ *and* $\lceil q\alpha'\rceil+\lceil q\beta'\rceil=1$.

The proof of this theorem can be found in literature [17]. The theorems presented in the next part of this paper are based on the above theorems. The presented methods find application in signal processing [18].

### 3.1. Data model

The data model $(s,\tau)$ presented in literature [19,4,20] makes it possible to describe the phenomena that contain various states at the same moment in time. This is an excess solution for the purposes of describing a data stream, where two various tuples do not exist at the same moment in time. Such an assumption significantly complicates the definition of the introduced operators [21]. The data stream for the needs of a data stream management system is described [7,22] by an ordered pair $(s_n,\Delta_n)$ whose first element is a sequence of tuples and the second one – the sequence of time intervals between the time occurrence of tuples. The data model in the form of $(s_n,\tau_n)$ describes the same class of phenomena as the model $(s,\tau)$. However, a differential model $(s_n,\Delta_n)$ prevents the presentation of a data stream where two different tuples appear at the same moment in time $\tau$.

When narrowing the range of the model's application, we assumed [7] that the sequence of time intervals between consecutive tuples is constant. As a result, this data model may be described in the following way: $(s_n, \Delta)$. This data model corresponding to a time series provides a basis for further deliberations.

It is assumed that relations are marked by Latin alphabet letters [23]. Similar notation for data streams has been accepted. Additionally, a constant value is linked with each data stream. This value denotes a constant interval between consecutive tuples. This value is noted after the name of the stream and after a comma, e.g. A,3 denotes stream A whose tuples come in every three seconds. Each stream contains a schema. A stream schema is a free, ordered set of attributes describing consecutive fields of tuples which belong to a data stream. By analogy with a relational model, each attribute corresponds to a data column. For example, the identifier of stream A(name,forename),3 denotes stream A whose each tuple contains two attributes: name and forename and each consecutive tuple appears every three seconds. By a uniform schema we understand a schema where all attributes are of the same type.

### 3.2. Introduced operators

After an analysis of most typical operations on data received from medical devices, the following set of operations was identified: Interlace, Deinterlace, Sum, Difference, Projection, Selection, Aggregation and Serialization (AGSE) and Offset. Assuming that there are two streams: A and B, these operations were defined in the following way:

**Sum and difference.** In the case of joining data streams obtained from devices that work synchronously, the streams must be joined in a sort of natural way, by joining their schemas. While the problem of summary joining is not difficult in data streams whose incoming elements appear with the same frequency, the elements of the slower stream must be duplicated in data streams whose elements come in with different frequency.

When considering the differences between a set of relations and a data stream set, the operation of summary joining (sum) and differential disjoining (difference) was defined in the following way:

The value of expression $\Sigma(A,B)$ is a data stream. The method of creating successive output stream tuples and determining value $\Delta$ of this stream is denoted by means of the following formula:

$$c_n = \begin{cases} a_n \,|\, b_{\left\lfloor \frac{n\Delta_a}{\Delta_b} \right\rfloor} & \Delta_c = \Delta_a \\ a_{\left\lfloor \frac{n\Delta_b}{\Delta_a} \right\rfloor} \,|\, b_n & \Delta_c = \Delta_b \end{cases}, \quad \Delta_c = \min\left(\Delta_a, \Delta_b\right). \tag{3}$$

Where:

$\Delta_{a,b,c}$ are the values that determine the constant time interval between tuples in streams $A$, $B$ and $C$,

$n$ denotes the position of a tuple,

$a_n$, $b_n$, $c_n$ denotes tuples of streams $A$, $B$ and $C$.

The integral part of the real number $x$ is marked by $\lfloor x \rfloor$. Ceiling dxe of the real number $x$ it is the lowest integral part not smaller than $x$.

**Example 1.** Assuming that there are exist two data streams:
Alfa(char),2: (1,2,3,4,...) and Beta(char),1: (a,b,c,d,e,f,g,...)
expression: $\Sigma(Alfa,Beta)$ describes the following data stream:
Omega(char,char),1: ((1,a),(1,b),(2,c),(2,d),...)

A difference operation enables the reconstruction of data streams that participate in a summary operation. It should be emphasized though that this is not a projection operation [1,23]. Tuples duplicated during the summary operation are removed during differential disjoining. Only the application of a projection operator enables the recovery of the primary form of data stream.

The argument of this operation is the data stream, (which has implicitly been summed) and a pair of two numbers that present a relation between time intervals of two primary data streams. The method of creating consecutive tuples of the data stream is denoted by the formula:

$$a_n = \begin{cases} c_n & \Delta_b \geq \Delta_a \\ c_{\left\lceil \frac{n\Delta_a}{\Delta_b} \right\rceil} & \Delta_b < \Delta_a \end{cases} \qquad (4)$$

The symbolic form of this operation takes the following form: $\delta_{a,b}(S)$. Where $a,b$ denote a pair of numbers that denotes a relation of time intervals between primary streams. $S$ is a stream – an argument of a difference operation.

Example 2. Assuming that there is a data stream Omega in the form presented above, then the expression $\delta_{a,b}(Omega)$ denotes a data stream in the form:
Alfa2(char,char),2: ((1,a),(2,c),(3,e),...)

**Interlace and deinterlace.** Summary interlace (Interlace) and differential deinterlace operations make an orthogonal set of operators in relation to summary and difference operations. While in the case of a sum and difference the stream schemes may be different, they must be in accordance with the schemes for interlace and deinterlace. Two schemes of data streams are in compliance if corresponding consecutive types of fields are the same.

If the frequency of both streams is the same, the output data stream after interlace contains tuples from stream arguments in turns. If the frequency of arguments is different, then the order of tuples is determined by the rule of interlace operation (5).

This operation is denoted in the following way: $\phi(A,B)$. The following formula presents the method of determining consecutive tuples of the output stream.

$$c_n = \begin{cases} b_{n-\lfloor nz \rfloor} & \lfloor nz \rfloor = \lfloor (n+1)z \rfloor \\ a_{\lfloor nz \rfloor} & \lfloor nz \rfloor \neq \lfloor (n+1)z \rfloor \end{cases}, \quad z = \frac{\Delta_b}{\Delta_a + \Delta_b}, \quad \Delta_c = \frac{\Delta_a \Delta_b}{\Delta_a + \Delta_b}. \tag{5}$$

**Theorem 3.** *The interlace operation assures a sequential cover of both sets containing elements of data streams.*

**Proof.** *In the first step, the first condition (equality) is analyzed in equation (5). For each n that fulfills condition* (*) $\lfloor nz \rfloor = \lfloor (n+1)z \rfloor$, *consecutive values of expression* $n - \lfloor nz \rfloor$ *create the second (corresponding) sequence of natural numbers that selects elements from sequence* $b_n$.

*This means that for each n that fulfills condition* (*) *the following relation:* $n - \lfloor nz \rfloor$ *should occur for the expression* $x_n = x_{n+1} - 1$. *The formal notation of this sentence is as follows:*

$$n - \lfloor nz \rfloor = (n+1) - \lfloor (n+1)z \rfloor - 1. \tag{6}$$

*Substituting condition* (*) *for the above formula, we obtain:*

$$n - \lfloor (n+1)z \rfloor = (n+1) - \lfloor (n+1)z \rfloor - 1. \tag{7}$$

*By a simple algebraic simplification we get to the following identity:* $n = (n + 1) - 1$

*The second part of the proof (based on the inequality condition) is analogous.* □

**Example 3.** If Alfa and Beta data streams exist, the expression $\phi$(Alfa,Beta) defines the interlace of two data streams that create the output stream in the form:

Gamma(char),2/3: (a,b,1,c,d,2,e,f,3,...)

However, expression $\phi$ (*Beta*,*Alfa*) defines the output stream in the form:

Delta(char),2/3: (1,a,b,2,c,d,3,e,f,...)

As we can see, the interlace is not a commutative operation.

Deinterlace is a complementary operation in comparison with the interlace operation. During stream deinterlace we can choose these tuples from the stream that belongs to one of the arguments of the interlaced stream. The arguments of this operation are a data stream and a number that defines the time interval. As a matter of form, two operators define this operation: $\Theta_n(A)$ and $\sim\Theta_n(A)$ The first one presents the operation of primary stream extraction and the second one – the residue from the deinterlace operation. Deinterlace is defined by the two formulas:

$$a_n = c_{n + \left\lceil \frac{(n+1)\Delta_a}{\Delta_b} \right\rceil}, \quad \Delta_a = \frac{\Delta_c \Delta_b}{|\Delta_c - \Delta_b|} \tag{8}$$

and

$$b_n = c_{n+\left\lfloor \frac{n\Delta_b}{\Delta_a} \right\rfloor}, \quad \Delta_b = \frac{\Delta_c \Delta_a}{\left|\Delta_c - \Delta_a\right|}. \tag{9}$$

**Theorem 4.** *The deinterlace operation is an instance of Fraenkel Partition Theorem.*

**Proof.** In the first part of the proof we look for the Beatty sequence (9) presented in the definition of the deinterlace operation. It is noted in the following way:

$$\left(n + \left\lfloor \frac{nb}{a} \right\rfloor\right)_{n=1}^{\infty}. \tag{10}$$

If $n \in \mathbb{N}$, then on the basis of property (21) the above equation might be presented as follows:

$$\left(\left\lfloor \frac{n-\alpha'}{\alpha} \right\rfloor\right)_{n=1}^{\infty} = \left(\left\lfloor n + \frac{nb}{a} \right\rfloor\right)_{n=1}^{\infty}. \tag{11}$$

For simplicity,

$$\left(\left\lfloor n\alpha^{-1} - \frac{\alpha'}{\alpha} \right\rfloor\right)_{n=1}^{\infty} = \left(\left\lfloor n\frac{a+b}{a} \right\rfloor\right)_{n=1}^{\infty}. \tag{12}$$

Symbol $-\alpha/\alpha'$ denotes an y-interception; if the value of sequence offset $\alpha' = 0$ then $\alpha = a/(a+b)$ which means that sequence (10) can be presented in the following way:

$$\mathcal{B}\left(\frac{a}{a+b}, 0\right) := \left(\left\lfloor n\frac{a+b}{a} \right\rfloor\right)_{n=1}^{\infty}. \tag{13}$$

The form of Beatty sequence is obtained through a few simple algebraic transformations from the sequence describing a deinterlace operation (9). The residue of sequence (10) is determined in the next part of the proof based on Fraenkel Partition Theorem.

1. The value of expression $\alpha = a/(a+b)$ for $a$, $b > 0$ is smaller than unity and greater than zero.
2. Condition $\alpha + \beta = 1$ is fulfilled for $\beta = b/(a+b)$.
3. For $\alpha' = 0$ postulate is equivalent to 1.
4. We search for solutions in a set of rational numbers.
5. If $q\alpha \in \mathbb{N}$ and $q \in \mathbb{N}$ and $1/q \le \alpha + \alpha'$, since $\alpha' = 0$ then $1/q \ge a/(a+b)$ and that is true for $q \le (a+b)/nwd(a,b)$.

From that there results $\left\lceil \frac{a+b}{nwd(a,b)} \beta' \right\rceil = 1$. That is $\beta' = \frac{nwd(a,b)}{a+b}$.

As a result of transformations, we obtain the form of the sequence presented (9) in deinterlace equation. The residue sequence form for $\mathcal{B}\left(a/\left(a+b\right),0\right)$ that fulfills the postulates of the Fraenkel Partition Theorem takes the following form:

$$\mathcal{B}\left(\frac{b}{a+b},\frac{nwd\left(a,b\right)}{a+b}\right)=\left(\left\lfloor\frac{\left(n+1\right)-\frac{nwd\left(a,b\right)}{a+b}}{\frac{b}{a+b}}\right\rfloor\right)_{n=1}^{\infty}. \quad (14)$$

By leaving out the brackets that describe the sequence, it can be demonstrated after a few simple transformations that:

$$\left\lfloor\frac{\left(n+1\right)-\frac{nwd\left(a,b\right)}{a+b}}{\frac{b}{a+b}}\right\rfloor:=\left\lfloor n\frac{a}{b}+n+\frac{a}{b}+1-\frac{nwd\left(a,b\right)}{b}\right\rfloor. \quad (15)$$

By comparing the determined form of the residue to the accepted sequence (8), we receive an equation the truth of which should be proved.

$$\left\lfloor n\frac{a}{b}+n+\frac{a}{b}+1-\frac{nwd\left(a,b\right)}{b}\right\rfloor=n+\left\lceil\frac{\left(n+1\right)a}{b}\right\rceil. \quad (16)$$

Hence, it can be shown that:

$$\left\lfloor\frac{\left(n+1\right)a}{b}-\frac{nwd\left(a,b\right)}{b}\right\rfloor+1=\left\lceil\frac{\left(n+1\right)a}{b}\right\rceil. \quad (17)$$

Substituting a natural number $n$ + 1 by $n$ we get:

$$\left\lfloor n\frac{a}{b}-\frac{nwd\left(a,b\right)}{b}\right\rfloor+1=\left\lceil n\frac{a}{b}\right\rceil. \quad (18)$$

In the further part of the proof, the following properties of ceiling and floor operations will be applied:

$$\lfloor x\rfloor=\lceil x\rceil\Leftrightarrow x\in\mathbb{N}, \quad (19)$$

$$\lfloor x\rfloor+1=\lceil x\rceil\Leftrightarrow x\in\mathbb{R}\backslash\mathbb{N}, \quad (20)$$

$$\lfloor x+C\rfloor=\lfloor x\rfloor+C\Leftrightarrow c\in\mathbb{N}. \quad (21)$$

The following properties of ceiling and floor and $nwd(a,b)$ operations will be applied further on:

$$nwd\left(a,b\right)=b\Leftrightarrow\frac{a}{b}=c\in\mathbb{N} \quad (22)$$

and

$$1 \le nwd(a,b) \le a \Leftrightarrow 0 < \frac{a}{b} < 1 . \tag{23}$$

Considering coincidence (22), equation (18) takes the form:

$$\left\lfloor \frac{(n+1)a}{b} - \frac{b}{a} \right\rfloor + 1 = \left\lceil \frac{(n+1)a}{b} \right\rceil . \tag{24}$$

Considering properties (19,21) and the domain for this coincidence (22), we can state that both sequences create the same elements.

Considering the next coincidence (23), it can be assumed that there exist two numbers *a* and *b* and that equation (18) is not false, so:

$$n\frac{a}{b} - \frac{nwd(a,b)}{b}, \quad n\frac{a}{b} \in \mathbb{N} \tag{25}$$

Using property (19,21) it is not true that:

$$\left\lfloor n\frac{a}{b} - \frac{nwd(a,b)}{b} + 1 \right\rfloor \ne \left\lceil n\frac{a}{b} \right\rceil . \tag{26}$$

Using property (19), we look for such *a* and *b* where the following property is true:

$$n\frac{a}{b} - \frac{nwd(a,b)}{b} + 1 \ne n\frac{a}{b} . \tag{27}$$

Equation (18) is fulfilled for $nwd(a,b) = b$, and in the considered domain (23) $1 \le nwd(a,b) \le a$ this equation does not have any solutions. There is no such *a* and *b* that belonging to domain (25) which negate considered equation (18).

After an analysis of the second property (20) we can assume that there are two such numbers: *a* and *b* that equation (18) is not fulfilled – this is for such *a* and *b* that:

$$n\frac{a}{b} - \frac{nwd(a,b)}{b}, \quad n\frac{a}{b} \in \mathbb{R} \setminus \mathbb{N} . \tag{28}$$

But there are no such two numbers which

$$n\frac{a}{b} - \frac{nwd(a,b)}{b} \ne n\frac{a}{b} . \tag{29}$$

But there do not exist two numbers which $nwd(a,b) = 0$

For $a/b \in \mathbb{R} \setminus \mathbb{N}$ this equation is always true.

Hence, both equations (8,9) are the instances of Beatty sequences that fulfill the postulates of the Fraenkel Partiton Theorem for rational numbers. □

**Example 4.** Assuming that the stream Gamma,2/3 presented in the previous example exists, the expression Gamma,2/3 then the expression $\Theta_1$(*Gamma*) describes the data stream:

Alfa(char),2: (1,2,3,4,...)

However expression $\Theta_2(Gamma)$ describes the data stream
Beta(char),1: (a,b,c,d,e,f,g,...)
**Projection.** A projection operator is used to create a new stream that arises from $A$ removing the columns or changing their order. The value of expression $\pi_{T_1,T_2,...,T_n,}(A)$ is a stream created from $A$ by copying values of all attributes $T_1, T_2, \ldots, T_n$. The projection operation does not affect the value of the constant time interval between tuples. The value of $\Delta$ is not changed after projection.
Example 5. If we assume the existence of the previously presented data stream Alfa2, then the expression $\pi_{Char}(Alfa2)$ describes a data stream in the form:
Alfa(char),2: (1,2,3,...)
**Selection.** After the selection operation on stream $A$ a new stream is created. This stream contains a subset of tuples from $A$. The tuples of the output stream are chosen according to the criteria described by condition $W$. This operation, ust like in relational algebra, is denoted by the symbol $\sigma_{w,\delta}(A)$. The output stream schema remains unchanged after selection. The order of attributes is also maintained. $W$ is a conditional expression that is false or true. The operands may be constants or stream $A$ attributes. If the expression is true, then the tuple is connected to the output stream, otherwise it is not. The selection modifies directly the constant value of time interval in the data stream. The value $\Delta$ is denoted from the top and this value is $\delta$.
**Example 6.** If we assume the existence of stream Omega(char,char),1 then the expression: $\sigma_{\text{is_number(char)},2}(Omega)$ describes the stream Alfa(char),2.
**Aggregation and serialization.** The acronym AGSE was created from the first syllables of the words Aggregation and Serialization. The argument of this operation is one data stream of uniform schema. After these operations, another data stream comes into being with a given schema and a modified, computed value of $\Delta$. This operation is denoted by the symbol $\Psi_{n,\lambda}(A)$. The symbol $\lambda$ means the step with which a sliding window is moved over the data stream. It is noted in the count of tuples. The symbol $n$ denotes the width of this window, which is expressed in the count of attributes of the schema.
**Example 7.** If there is stream Alfa(char),1 then the expression $\Psi_{2,2}(Alfa)$ denotes a new stream in the form: $Agse1(char,char)$,2 whose elements create the sequence: $((a,b), (c,d), (e,f),\ldots)$ . However, the expression $\Psi_{1,1}(Beta)$ denotes stream Beta(char),1. The first operation is an example of aggregation, the second one is serialization.
**Offset.** The offset operator is the main operator used in digital signal processing filters. Its role is to enable access to archival data. This operation creates a data stream whose elements are shifted in time by a given argument of the number of tuples. This operation was denoted by the $\tau_n(A)$ symbol. The symbol $n$ denotes the number of tuples by which the output stream should be delayed in relation to $A$.

**Example 8.** If stream Alfa(char),1 exists, then the expression: $\tau_2(Alfa)$ defines stream
Shift2(char),1: (3,4,5,6,...)

## 4. Properties of operations

The rule of moving the selection operations over joining operations is applied to query plans during time optimization [24]. Similar issues are considered in the case of stream systems [25]. In the course of research on a deterministic method of time optimization, some properties and algebraic methods of simplifying stream expressions were discovered. The first property concerns the disorder of sequence elements. Next, a relatively simple rule of summary commutative is described. The last method, described as the method of adjusting interlace, enables a change of the sequence of attributes in an interlacing operation if some additional conditions are fulfilled.

### 4.1. Disorder of the sequence of events

**Theorem 5.** *The sequence of elements in a stream does not always reflect the real sequence of elements in the real word.*
**Proof.** This property can be discovered in a simple way by analyzing an interlace operation used on the following data streams:
Alfa(char),2: (1,2,3,4,5,6...)
Epsilon(char),3: (a,b,c,d,e,f...)
The expression $\phi(Epsilon,Alfa)$ describes the interlace of data streams in the form:
Tau(char),6/5: (1,2,a,3,b,4,5,c,6,d,...)

In the Tau stream, the tuple denoted by letter c follows the tuple denoted by number 5. On the other hand, the tuple denoted by letter c occurs in second 9 in the Epsilon stream, and in the Alfa stream the tuple denoted by number 5 occurs in second 10. The natural order of events in output streams was broken. The conclusion is as follows: The sequence of tuples does not always correspond to the sequence in the real word. When carrying out an analysis of data in streams by time domain, this aspect should by emphasized and in such a case the operation of deinterlace should be used. This operation helps receive a primary form of data streams. □

### 4.2. Summary is commutative

**Theorem 6.** *A data stream summary operation is commutative without taking into account the sequence of attributes.*
Proof. We assume that $C = \Sigma(A,B)$ and $C = \Sigma(S,A)$. The following two relationships can be presented with the use of the data stream summary equation:

$$C = \Sigma(A,B) \quad c_n = \begin{cases} a_n | b_{\left\lfloor \frac{n\Delta_a}{\Delta_b} \right\rfloor} & \Delta_a < \Delta_b \\ a_{\left\lfloor \frac{n\Delta_b}{\Delta_a} \right\rfloor} | b_n & \Delta_a > \Delta_b \end{cases} \tag{30}$$

and

$$D = \Sigma(S,A) \quad d_n = \begin{cases} s_n | a_{\left\lfloor \frac{n\Delta_s}{\Delta_a} \right\rfloor} & \Delta_s < \Delta_a \\ s_{\left\lfloor \frac{n\Delta_a}{\Delta_s} \right\rfloor} | a_n & \Delta_s > \Delta_b \end{cases} \tag{31}$$

Without taking into account the sequence of attributes that result from the junction operation of elements of tuples and if the sequence of conditions is changed, the following formula can be presented:

$$D = \Sigma(S,A) \quad d_n = \begin{cases} a_n | s_{\left\lfloor \frac{n\Delta_a}{\Delta_s} \right\rfloor} & \Delta_s > \Delta_a \equiv (\Delta_a < \Delta_s) \\ a_{\left\lfloor \frac{n\Delta_s}{\Delta_a} \right\rfloor} | s_n & \Delta_s < \Delta_b \equiv (\Delta_a > \Delta_s) \end{cases} \tag{32}$$

By substituting symbol $S$ in formula (32) by symbol $B$, we prove the correctness of the thesis about the commutation of data stream summary. There is one more case in which $\Delta$ of both data streams is equal. The proof was not presented due to triviality. □

### 4.3. Interlace adjusting method

As presented in the example of interlace operation, interlace is not commutative. However, there is an algebraic method that enables the change of the sequence of attributes under some conditions.

**Theorem 7.** *If we choose two natural numbers i,k whose relation is equal to the value of relation of $\Delta_n$ elements joined by interlace data streams, then the operation of interlace shifted by these values creates a data stream which is equal to a stream created by interlace and shifted by the sum of these two numbers and change of the sequence of arguments of this operation.*

*Formally: if the following conditions are fulfilled:* $i/k = \Delta_a / \Delta_b$ *and* $i,k \in \mathbb{N}$ *then:*

$$\phi\left(\tau_i(A), \tau_k(B)\right) = \tau_{i+k}\left(\phi(B,A)\right). \tag{33}$$

**Proof.** By analyzing the left side of equation (33) and considering an equation that describes interlace operation (5), the following formula can be reached:

$$C=\phi\left(\tau_i\left(A\right),\tau_k\left(B\right)\right)\quad c_n=\begin{cases} b_{\left(n-\lfloor nz\rfloor\right)+i} & \lfloor nz\rfloor=\lfloor\left(n+1\right)z\rfloor \\ a_{\left(\lfloor nz\rfloor\right)+k} & \lfloor nz\rfloor\neq\lfloor\left(n+1\right)z\rfloor \end{cases}. \tag{34}$$

By analyzing the right side of equation (33) and considering an equation that describes interlace operation (5), the following formula can be reached:

$$C=\tau_{i+k}\left(\phi\left(B,A\right)\right)\quad c_n=\begin{cases} a_{\lfloor\left(n+i+k\right)z\rfloor} & \lfloor nz\rfloor=\lfloor\left(n+1\right)z\rfloor \\ b_{n+i+k-\lfloor\left(n+i+k\right)z\rfloor} & \lfloor nz\rfloor\neq\lfloor\left(n+1\right)z\rfloor \end{cases}. \tag{35}$$

Considering conditions for which equations choose proper samples from streams *A*, *B* and using the assumptions of theorem (33), we can state that:

$$b_{n-\lfloor nz\rfloor+i}=b_{n+i+k-\lfloor\left(n+i+k\right)z\rfloor}\Rightarrow-\lfloor nz\rfloor=k-\lfloor\left(n+i+k\right)z\rfloor \tag{36}$$

and

$$i+k=\frac{\Delta_a}{\Delta_b}k+k=k\left(\frac{\Delta_a}{\Delta_b}+1\right)=\frac{k}{z}. \tag{37}$$

Given the assumptions of thesis (33) and the previous equations, the following conclusion can be drawn:

$$-\lfloor nz\rfloor=k-\lfloor k+nz\rfloor. \tag{38}$$

Taking into consideration the assumption of thesis (33), i.e. $k\in\mathbb{N}$ it can be stated that equation (38) is fulfilled. The second part of the proof based on the inequality condition is analogous. □

## 5. Construction of query plans

A set of operators and methods enabling improvement of effectiveness of data access was presented in the previous sections. The effectiveness of realization is managed by a query processor. The query processor is an element of a database management system that serves as a translator of queries expressed in a formal language into the sequence of operations on a data set. The wrong strategy of query realization may lead to algorithms which process queries in a considerably longer period of time than necessary. In the majority of relational database management systems, the internal form of SQL queries is expressed in the form of relational algebra. An advantage of relational algebra is the possibility of presenting algebraic expressions in the form of expression trees that present a logical query plan [24].

Similar principles were accepted for the query processor of the created system. Stream algebra is used in order to express the internal query form presented in this paper. The defined set of operators for this system is different from relational algebra operators although, similarly, expression trees were used. A sequence of phases is executed by a query processor to build an expression tree [24]. First, a query processor checks up on the syntax of the issued query. A query language is described through formal grammars – in particular context free grammar [26]. These grammars allow for a clear representation of the query structure in the form of a tree expression and enable the creation of an algorithm that will accept grammatically correct expressions. Next, the expression tree should be transformed into a tree of algebraic operators according to the algebra used. In this phase, the database management system decides which algorithm is the best for the realized task. The created expression tree is also called a logical query plan. In the last phase, the logical query plan is transformed into a physical query plan. This query plan contains the executed physical operations in the proper sequence, as well as the method of flow control in this algorithm.

### 5.1. Expressions tree

One algebraic expression may contain several operators. They work like a cascade on the results of previous operations. Therefore, the sequence of the execution of operators may be presented as a tree of expressions. The leaves are the stream names, each bend tag is an operator.

Let us assume that there are the two data streams: ECG( patient_id, ecg_v ), 0.001 and Sensor( sensor_id, value ),0.1. The first data stream contains information about the measured electrical heart potentials, while the second stream contains the measure from the biomedical sensor. The query which creates the stream containing ECG probes whose value of sensor is greater than 25 is following:

```
SELECT ecg_v
STREAM bol_ekg
FROM ECG + Sensor
FILTER BY value > 25, 0.001
```

Such a formalized query is translated by a query analyzer into a logical query plan. In the first step, the streams ECG and Sensor are joined. In the next step there is selection corresponding to the FILTER BY clause. The last step is the projection to attributes contained in the SELECT. The algebraic expression of the above query is presented as a tree in Figure 1.

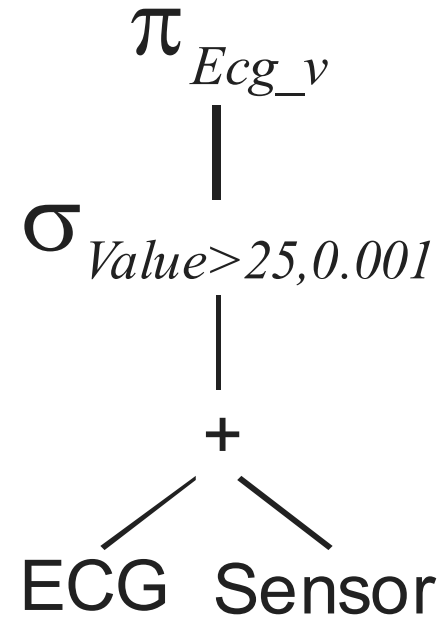

Fig. 1. Logical query plan

There are a few expressions equivalent to the one presented in Figure 1. in such a way that various query plans generate the same results. A general principle states that selection should (usually) be executed as soon as possible. A similar rule applies to data stream processing because other operations usually take much longer than selection which reduces the size of data streams.

The use of the introduced method of interlace adjustment (33) by the query processor and commutative of the summary (30,31) permits the selection of a more appropriate query plan.

## 6. Conclusions

Within the framework of scientific research, tasks are realized with the aim of creating monitoring systems for medical purposes [6]. In most cases, the data management task is realized by a monitoring system. In an early stage such a solution does not complicate significantly the construction of the system. The situation changes afterwards. It turns out that the already implemented algorithms are strongly connected with the chosen implementation mode and therefore it is difficult to use the latest hardware solutions. This work aims at presenting a data management system with a formal query language as a solution to this problem. The solution is based on the concept of stream query processing. The application of the proposed system simplifies the construction of digital signal processing algorithms, assures data safety and offers a solution to multi-usage of information in a monitoring system.

Data stream management systems are currently one of the most intensively developed fields of science connected with database systems. The aim of this work is to prove the efficiency of the presented data processing model with relation to the existing solutions [27]. This task is rather difficult because the accepted relational model is a universal one. The application of the stream data model is expected to be simpler and more elegant in a monitoring system. Thanks to a simplified construction, the effectiveness of data processing will grow significantly. Yet, the level of universality of the relational model will be difficult to achieve.

This paper presents the theorems of data sequence (streams) algebra including proofs developed in scientific research. A direct connection between the used operators and the Theory Numbers Theorems has been proved. These theorems were proved again a relatively short time ago [17].